\documentclass[A4paper,preprint,12pt,tightenlines,amsmath,amssymb,floatfix,nofootinbib,aps,raggedbottom]{revtex4-2}

\usepackage{graphicx}
\usepackage{bm}
\usepackage{url,multirow,xcolor}
\usepackage[font=footnotesize,margin=24pt,justification=centerlast]{caption}
\usepackage{adjustbox}

\begin{document}

\title{ A  solution to the uncertainty problem of the nuclear matrix element for neutrinoless double-$\bm{\beta}$ decay.
\vspace{5pt} }

\author{J.\ Terasaki \\ \vspace{0pt}}
\affiliation{ Institute of Experimental and Applied Physics\hbox{,} Czech Technical University in Prague, Husova 240/5, 110\hspace{3pt}00 Prague 1, Czech Republic}

\author{O.\ Civitarese \\ \vspace{0pt}}
\affiliation{\raisebox{0pt}{Department of Physics, University of  La Plata, 49 y 115.~C.C.~67 (1900)}, La Plata, Argentina \\
and \hbox{IFLP-CONICET, diag 115 y 64. La Plata, Argentina}\vspace{10pt}}

\begin{abstract} 
The neutrinoless double-$\beta$ decay ($0\nu\beta\beta$) of nuclei is one of the major research subjects of neutrino physics nowadays because of its influence on particle physics and astrophysics. 
The predicted nuclear matrix elements (NMEs) for the $0\nu\beta\beta$ decay exhibit large uncertainties depending on the models employed. This problem has affected the development of neutrino physics for many years. We have recently performed the calculation of the NMEs for  the $0\nu\beta\beta$  and two-neutrino double-$\beta$ decay ($2\nu\beta\beta$)  modes  with a perturbed transition operator and found that the effective axial-vector current coupling $g_A^\mathrm{eff}$ is similar for the two decay modes. Based on this finding, we calculate the $0\nu\beta\beta$ NMEs using the phenomenological $g_A^\mathrm{eff}$ that reproduces the measured half-life of the $2\nu\beta\beta$ decay. We apply this method to the NMEs for $^{136}$Xe, $^{130}$Te, and $^{76}$Ge obtained by several groups and show that the uncertainty of the $0\nu\beta\beta$ NME is dramatically reduced. Based on this result, we calculate the effective neutrino mass, consistent with the  current experimental lower limit of the half-life for the  $0\nu\beta\beta$ decay. The results indicate that the extracted  value of the effective neutrino mass does not yet reach the inverted-mass-hierarchy region allowed by neutrino oscillation data when the lightest neutrino mass is assumed to be below 10 meV. We also calculate the perturbed $0\nu\beta\beta$ and $2\nu\beta\beta$ NMEs of $^{110}$Pd. 
\end{abstract}


%
\maketitle
\newpage
\section{\label{sec:introduction}Introduction}
The neutrinoless double-$\beta$ ($0\nu\beta\beta$) decay is a door to new physics because its detection would imply the need to extend the standard model of the electroweak interactions and will establish the Majorana nature of the neutrino. 
This decay is expected to occur from one nucleus to another with a suitable Q value. Thus, the study of the $0\nu\beta\beta$ decay is an interdisciplinary field involving particle physics and nuclear physics. 
The nuclei expected to have the $0\nu\beta\beta$ decay are relatively heavy ones, so that approximations are necessary to calculate the nuclear part of the transition matrix element (NMEs). Nuclear theorists have devoted intensive effort to obtain the accurate NMEs for more than 40 years, but the calculated NMEs are distributed in a broad range \cite{Ago23}.

The calculations cannot be confirmed directly by experiments, and the uncertainty in the values of the NMEs  has not been reduced since this problem was recognized. The purpose of this article is to show a possible solution to this uncertainty problem affecting  the $0\nu\beta\beta$ NMEs. 

Recently, we have extended the equation of the NME for the $0\nu\beta\beta$ decay \cite{Ter25} in terms of the perturbed transition operator. The usual equation, which we call the leading order one, is derived using the first-order Rayleigh-Schr\"{o}dinger (RS) perturbation method. We derived the NME including the transition operator perturbed by the nucleon-nucleon interaction, using the second-order RS perturbation and the diagrams expressing the decay process. Our higher-order terms consist of the vertex correction and the two-body current terms, which we call RS two-body current term to distinguish it from that of the chiral effective field theory ($\chi$EFT). The corresponding equations for the two-neutrino double-$\beta$ ($2\nu\beta\beta$) NME were also derived. 

By applying this scheme we have obtained the effective value of the axial-vector current coupling $g_A$ ($g_A^\mathrm{eff}$) to be used with the leading-order components of the NMEs, referring to the perturbed NMEs with the bare value of $g_A$ for the nucleon in vacuum ($g_A^\mathrm{bare}$). We obtained this theoretical $g_A^\mathrm{eff}$ for the $0\nu\beta\beta$ and $2\nu\beta\beta$ decays. 

This effective parameter $g_A^\mathrm{eff}$ contains the many-body effects that are not fully included in the leading-order NMEs. It is shown that the theoretical values of $g_A^\mathrm{eff}$'s for the two decay modes are close. Since $g_A^\mathrm{eff}$ for the $2\nu\beta\beta$ decay is known experimentally, we apply this $g_A^\mathrm{eff}$ to the $0\nu\beta\beta$ NME and explore the consequences of this approach upon the uncertainty of the NMEs. Another possible approach is to keep $g_A^\mathrm{bare}$ and calculate the NME including higher-order perturbations \cite{Men11}. 

The mathematical details of our method are explained in Ref.~\cite{Ter25}. The use of it is described in Sec.~\ref{sec:method}, for the sake of completeness.  We apply our method for the decays of $^{136}$Xe in Sec.~\ref{sec:xe136} and of $^{76}$Ge in Sec.~\ref{sec:ge76}. It is shown that the dispersion of the NMEs, as provided from the published works of several groups, is reduced significantly, and that the NME values are lowered. The consequences of these results upon the estimation of the effective neutrino mass are discussed at the end of Sec.~\ref{sec:ge76}. We also show the  perturbed NME for the $\beta\beta$ decays of $^{130}$Te and $^{110}$Pd in Sec.~\ref{sec:130Te_110Pd}. 
The main results of the present work are summarized in Sec.~\ref{sec:summary}. 

\section{\label{sec:method}Our method}
The technical framework of our calculation can be summarized below.
The perturbed NMEs for the $0\nu\beta\beta$ and the $2\nu\beta\beta$ decays of $^{136}$Xe, $^{130}$Te, $^{110}$Pd, and $^{76}$Ge are calculated using the quasiparticle random-phase approximation (QRPA) in conjunction with the Skyrme and the Coulomb interactions for the particle-hole interactions and the contact pairing interactions for the particle-particle interactions. The strengths of the like-particle pairing interactions are determined to reproduce the empirical pairing gaps obtained from the odd-even mass differences through the three-point mass formula \cite{Boh69} by the Hartree-Fock-Bogoliubov (HFB) calculations \cite{Ter03, Bla05, Obe07} of the ground states. 
The strength of the isovector proton-neutron pairing interaction is fixed at the average value of the proton and neutron pairing strengths.
This approximate isospin invariance leads to the almost vanishing  
Fermi component of the $2\nu\beta\beta$ NMEs. The strength of the isoscalar proton-neutron pairing interaction is determined by the method developed in Refs.~\cite{Ter16, Ter18}, based on the equality of the $0\nu\beta\beta$ NME to that obtained in the process of the two-neutron annihilation followed by the two-proton addition under the closure approximation. The convergence of the results with respect to the size of the single-particle basis and the number of the excited states of the participant double-odd mass nucleus was checked.

 We use the perturbed Gamow-Teller (GT) NME ($M_{0\nu}^\mathrm{GT}$) and  Fermi NME ($M_{0\nu}^\mathrm{F}$) for the $0\nu\beta\beta$ decay and $M_{2\nu}^\mathrm{GT}$ and $M_{2\nu}^\mathrm{F}$ for $2\nu\beta\beta$ decay. The leading-order NMEs are denoted by $M_{0\nu}^\mathrm{GT(0)}$, $M_{0\nu}^\mathrm{F(0)}$, $M_{2\nu}^\mathrm{GT(0)}$, and $M_{2\nu}^\mathrm{F(0)}$, respectively. By taking the phase space factors and the effective neutrino mass (Majorana neutrino mass) as constants and taking the value $g_V$ = 1 for the vector current coupling, the half-life for the $0\nu\beta\beta$ decay $T_{1/2}^{0\nu}$ and that for the $2\nu\beta\beta$ decay $T_{1/2}^\mathrm{2\nu}$ can be written as a function of the NMEs and $g_A$ \cite{Ter25}. 

The theoretical values of $g_A^\mathrm{eff}$ for the leading-order NMEs  for the two decay modes  can be determined by 
\begin{eqnarray}
T_{1/2}^\mathrm{0\nu}(M_{0\nu}^\mathrm{GT(0)}, M_{0\nu}^\mathrm{F(0)}, g_{A,0\nu}^\mathrm{eff}(\mathrm{ld;pt})) = T_{1/2}^\mathrm{0\nu}(M_{0\nu}^\mathrm{GT}, M_{0\nu}^\mathrm{F}, g_{A}^\mathrm{bare}), \label{eq:cndn_gaeff0v}\\
T_{1/2}^\mathrm{2\nu}(M_{2\nu}^\mathrm{GT(0)}, M_{2\nu}^\mathrm{F(0)}, g_{A,2\nu}^\mathrm{eff}(\mathrm{ld;pt})) = T_{1/2}^\mathrm{2\nu}(M_{2\nu}^\mathrm{GT}, M_{2\nu}^\mathrm{F}, g_{A}^\mathrm{bare}). \label{eq:cndn_gaeff2v}
\end{eqnarray}
The argument ld of $g_A^\mathrm{eff}$ indicates that this $g_A^\mathrm{eff}$ is used with the leading-order NME components, and the argument pt indicates that the perturbed NME components are referred to.

In our previous study for $^{136}$Xe \cite{Ter25}, we found numerically the relation 
\begin{eqnarray}
g_{A,0\nu}^\mathrm{eff}(\mathrm{ld;pt}) = 1.14\,g_{A,2\nu}^\mathrm{eff}(\mathrm{ld;pt}), \label{eq:relation_g0veff_g2veff}
\end{eqnarray}
with the Skyrme energy density functional (interaction) SkM$^\ast$ \cite{bar82}. The leading-order relation is trivially 
\begin{eqnarray}
g_{A,0\nu}^\mathrm{eff}(\mathrm{ld;ld}) = g_{A,2\nu}^\mathrm{eff}(\mathrm{ld;ld}) = g_A^\mathrm{bare}.
\label{eq:gA0veffldest}
\end{eqnarray}
Thus, Eq.~(\ref{eq:relation_g0veff_g2veff}) implies that this similarity relation between $g_{A,0\nu}^\mathrm{eff}$ and $g_{A,2\nu}^\mathrm{eff}$ is close to the convergence at the first order. 

It is known that the momentum-dependent form factors for the weak-interaction vertex only cause small modifications to the $0\nu\beta\beta$ NME \cite{Sim08}. Thus, those form factors do not cause a significant difference between $g_{A,0\nu}^\mathrm{eff}$ and $g_{A,2\nu}^\mathrm{eff}$. The effect of the chiral two-body current on the $0\nu\beta\beta$ and the $2\nu\beta\beta$ NMEs was investigated in Ref.~\cite{Eng14}. 
Taking the relevant NME values from Fig.~2 in this reference (the nucleus is $^{76}$Ge) and applying our procedure, we can derive the equation  
\begin{eqnarray}
g_{A}^\mathrm{eff}(0\nu\beta\beta) = 1.10\hspace{1pt} g_A^\mathrm{eff}(2\nu\beta\beta), \label{eq:relation_g0veff_g2veff_Engel}
\end{eqnarray}
where $g_{A}^\mathrm{eff}(0\nu\beta\beta)$ and $g_{A}^\mathrm{eff}(2\nu\beta\beta)$ correspond to $g_{A,0\nu}^\mathrm{eff}(\mathrm{ld;pt})$ and $g_{A,2\nu}^\mathrm{eff}(\mathrm{ld;pt})$, respectively. This equation obtained independently is consistent with  Eq.~(\ref{eq:relation_g0veff_g2veff}).

The theoretical reason for this similarity was already discussed in Ref.~\cite{Ter25}, and it stated
that $g_A^\mathrm{eff}$ is given by the ratio between  the perturbed NME component and the leading-order NME component. 
For $g_{A,0\nu}^\mathrm{eff}$, the neutrino potential is included in both the numerator and the denominator of that ratio, then it is approximately cancelled out; see Ref.~\cite{Ter25} for mathematical justification. Therefore the similarity of $g_{A,0\nu}^\mathrm{eff}$ and $g_{A,2\nu}^\mathrm{eff}$ is not just a coincidence. 

This similarity allows us to approximate $g_{A,0\nu}^\mathrm{eff}$ by the value of $g_{A,2\nu}^\mathrm{eff}$ which reproduces the experimental $2\nu\beta\beta$ half-life $T_{1/2}^{2\nu\mathrm{(exp)}}$ \cite{Bar19}. This value of $g_{A,2\nu}^\mathrm{eff}$ for the leading-order NME is denoted by $g_{A,2\nu}^\mathrm{eff}$(ld;exp). The non-perturbative effect on the NMEs can be taken into account by using the experimental data. For the application, we use the expression \cite{Ter25b}
\begin{eqnarray}
g_{A,0\nu}^\mathrm{eff}(\mathrm{ld;est}) = g_{A,2\nu}^\mathrm{eff}(\mathrm{ld;exp})\frac{g_{A,0\nu}^\mathrm{eff}(\mathrm{ld;pt})}{g_{A,2\nu}^\mathrm{eff}(\mathrm{ld;pt})}, \label{eq:gA0veff_est}
\end{eqnarray}
which we call the estimated $g_{A,0\nu}^\mathrm{eff}$. The fraction in the right-hand side is called the renormalization factor. 
If the perturbation gives the exact NME, the right-hand side is equal to the exact $g_{A,0\nu}^\mathrm{eff}$. We use $g_{A,0\nu}^\mathrm{eff}(\mathrm{ld;est})$ and the leading-order $0\nu\beta\beta$ NME components to obtain the effective NME. 

\begin{figure}[t] 
\includegraphics[width=0.5\columnwidth]{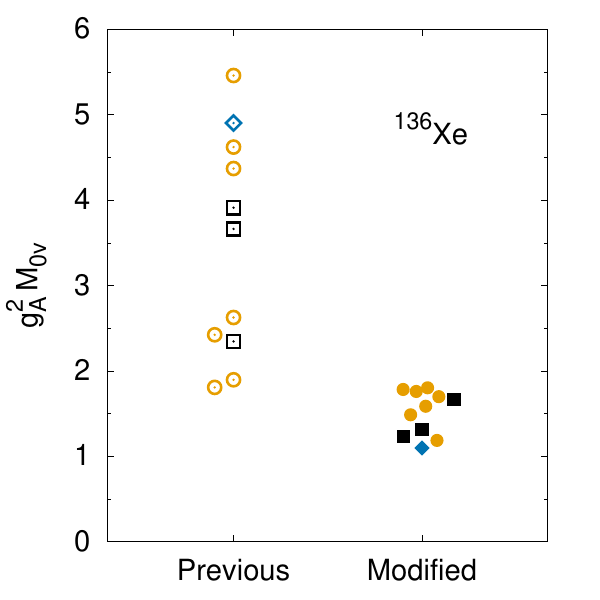}
\vspace{-10pt}
\caption{\label{fig:gA2M0v} \baselineskip=13pt 
The $0\nu\beta\beta$ NMEs multiplied by $g_A^2$ for $^{136}$Xe. The open symbols are the values   given by Eq.~(\ref{eq:gA2M0v_previous}). The filled symbols are the results  given by Eq.~(\ref{eq:gA2M0v_modified}). The  NMEs are shown  with circles (QRPA \cite{Mus13, Hyv15, Sim18b, Fan18, Ter25} and our calculation), squares (Shell Model \cite{Hor13, Men18}), 
 and diamonds (IBA \cite{Bar15}.)}
\end{figure}

Now, this procedure can be applied for any $0\nu\beta\beta$ calculations, as long as the leading-order $0\nu\beta\beta $ and $2\nu\beta\beta$ NMEs are calculated using the same interaction, the same nuclear wave functions, and the same approximation for the NME except for the closure approximation.

\section{\label{sec:xe136}The decay of $^{\mathbf{136}}$X\lowercase{e} }
The first example is the $\beta\beta$ decay of $^{136}$Xe, for which we have already tested our method   with NMEs calculated within the framework of the QRPA and using two types of the Skyrme interaction \cite{Ter25,Ter25b}.

We have selected the NMEs calculated by several groups, satisfying the condition for our procedure, and a set of eleven NMEs is used. This set consist of the NMEs obtained with the QRPA \cite{Mus13, Hyv15, Sim18b, Fan18, Ter25}, the Shell Model \cite{Hor13, Men18} and the Interacting Boson Approximation (IBA) \cite{Bar15} methods. We have applied our renormalization factor for $g_{A,0\nu}^\mathrm{eff}(\mathrm{ld;est})$ for these other calculations. When our procedure is applied for different calculations, $g_{A,0\nu}^\mathrm{eff}(\mathrm{ld;est})$ depends on the calculation. Thus, we use 
\begin{eqnarray}
[g_{A,0\nu}^\mathrm{eff}\mathrm{(ld;est)}]^2 M_{0\nu}^\mathrm{eff} =
[g_{A,0\nu}^\mathrm{eff}\mathrm{(ld;est)}]^2 M_{0\nu}^\mathrm{GT(0)} - g_V^2 M_{0\nu}^\mathrm{F(0)}, 
\label{eq:gA2M0v_modified}
\end{eqnarray}
for comparison\footnote{The tensor term of the NMEs would be added if it is also included in the transition operators. The contributions from the right-handed neutrinos are not considered. The NMEs of other groups not including the vertex correction or the two-body current term \cite{Ter25} are taken as the leading-order NMEs in our terminology. The NMEs including similar correction terms are distinguished from the leading-order NMEs.}. This is the nuclear factor affecting $T_{1/2}^{0\nu}$.
The corresponding expression with these  NMEs, before the modifications, is written 
\begin{eqnarray}
(g_A^\mathrm{bare})^2 M_{0\nu}^\mathrm{(0)} =
(g_A^\mathrm{bare})^2 M_{0\nu}^\mathrm{GT(0)}- g_V^2 M_{0\nu}^\mathrm{F(0)},
\label{eq:gA2M0v_previous}
\end{eqnarray}
where $g_A^\mathrm{bare}$ = 1.25$-$1.27 depending on the group. 

The comparison of $(g_A^\mathrm{bare})^2 M_{0\nu}^{(0)}$  and $[g_{A,0\nu}^\mathrm{eff}(\mathrm{ld;est})]^2 M_{0\nu}^\mathrm{eff}$  is shown in Fig.~\ref{fig:gA2M0v}. 
The former quantity is called previous, and the latter is called modified below. 
The range of values of the modified NMEs is reduced to one-fourth of the interval of the original ones. 
The significant implication of this figure is that the dispersion of the NME can be reduced, and actually many calculations indicate the low values. 

Here, we summarize the features of the different theoretical methods with reference to our treatment. Horoi and Brown \cite{Hor13} (Shell Model) introduced a quenching factor $q = 0.74 $ to the GT operator for the $2\nu\beta\beta$ decay. They concluded that, in this manner with their largest single-particle space, the $2\nu\beta\beta$ experimental data are reproduced.
We assume that they did not use this quenching factor $q$ for the $0\nu\beta\beta$ NME, and we multiply $(g_A^\mathrm{bare} q \times 1.14)^2$ to their total $0\nu\beta\beta$ NME, in order to be consistent with the requirements of our modification method. The $0\nu\beta\beta$ NMEs are given in their article. 

Men\'{e}ndez \cite{Men18} (Shell Model) used two short-range correlations (SRC), so that we use their two NMEs. The SRC is a function to approximate the high-momentum component of the $0\nu\beta\beta$ NME. The $2\nu\beta\beta$ NME is given in Ref.~\cite{Cau12}. They did not use a quenching factor or an effective $g_A$. 

Mustonen and Engel \cite{Mus13} (QRPA) used $g_A$ = 1.27 and 1.0 for the $0\nu\beta\beta$ NME, and the strength of the isoscalar pairing interaction was determined to reproduce the experimental $2\nu\beta\beta$ NME for these two values of $g_A$. We use their result with $g_A$ = 1.27 for the previous one and that with $g_A$ = 1.0 for the modified one. The latter is physically more reliable than the former because the larger $g_A$ leads to results close to the breaking point of the QRPA. They did not show the NME components. They used the Skyrme interaction SkM$^\ast$ and the modified SkM$^\ast$ \cite{Mus13}, so that we use the two samples for each of the previous and modified NMEs. 

Fang et al.~\cite{Fan18} (QRPA) used two $g_A$'s ($g_A^\mathrm{bare}$ and $0.75\hspace{1pt}g_A^\mathrm{bare}$) and  two interactions of AV18 and CD Bonn with two SRCs. The two samples that give the maximum and the minimum $0\nu\beta\beta$ NMEs with $g_A^\mathrm{bare}$ are used for the NMEs before applying our modification, and their corresponding calculations with $0.75\hspace{1pt}g_A^\mathrm{bare}$ are used for the modified one. 

The QRPA approach uses the overlap of the two intermediate states obtained on the basis of the initial and the final ground states. Fang et al.~used an approximate equation of the QRPA intermediate-state overlap proportional to the BCS ground state overlap of the initial and the final nuclei. This property is exact, as seen from the exact equation of the QRPA intermediate-state overlap \cite{Ter13}. They have obtained the BCS ground-state overlaps of 0.43 (AV18 interaction) and 0.39 (CD Bonn potential). Our calculations use the HFB ground states, and our HFB overlap value is 0.38. That of Mustonen and Engel is 0.49. Thus, these independent calculations give similar values. 
If the overlap of two normalized wave functions is equal to 1, the two wave functions are identical. If the particle number is a good quantum number, the overlap of the wave functions of different nuclei vanishes. Therefore small values of the  BCS overlap are more reliable than those close to 1. 
Fang et al.~emphasize that the NME is reduced due to the small BCS ground-state overlap. 
We speculate that their previous calculations (the spherical QRPA) do not include the BCS ground-state overlap (see Ref.~\cite{Sim04}). 

\v{S}imkovic et al.~\cite{Sim18b} (QRPA) did not mention the BCS ground-state overlap in their article. To be consistent with the other modified NMEs, we multiplied the BCS ground-state overlap of Fang.~et al.~of 0.43 (AV18 interaction) to the modified NME of \v{S}imkovic et al. 
They used $g_A^\mathrm{bare}$ for their $0\nu\beta\beta$ calculation and determined the strengths of the isovector and the isoscalar pairing interactions, proportional to the particle-hole interaction, so as to have the $2\nu\beta\beta$ GT and Fermi NMEs to vanish under the closure approximation. Their $2\nu\beta\beta$ Fermi NME without the closure approximation is not negligible for $^{136}$Xe. To avoid an ambiguity, we did not take their Fermi component into account.
For the modified result of Hyv\"{a}rinen and Suhonen \cite{Hyv15}, we have included also the BCS overlap factor. 
The $0\nu\beta\beta$ NME of Barea et al.~(IBA) was taken from Ref.~\cite{Bar15}, in which $g_A^\mathrm{bare}$ is used. The $2\nu\beta\beta$ NME was taken from Ref.~\cite{Bar13ibm}.

Let us now  discuss the value of the effective neutrino mass $\langle m_\nu \rangle$ which can be extracted  from the half-life $T_{1/2}^{0\nu}$. The relation between them is given by the decay probability associated with the half-life
\begin{eqnarray}
\frac{1}{T_{1/2}^{0\nu}} = G_{0\nu}|g_A^2 M_{0\nu}|^2\left(\frac{\langle m_\nu\rangle}{m_e}\right)^2.
\end{eqnarray}
Here, $G_{0\nu}$ is the phase space factor for the $0\nu\beta\beta$ decay \cite{Kot12}, not including $g_A^4$, and 
$g_A^2 M_{0\nu}$ represents either $[g_{A,0\nu}^\mathrm{eff}(\mathrm{ld;est})]^2 M_{0\nu}^\mathrm{eff}$ or $(g_A^\mathrm{bare})^2 M_{0\nu}^{(0)}$ in this article. The electron mass is denoted by $m_e$. Currently, the value of $\langle m_\nu \rangle$ cannot be fixed  by its definition in terms of the Pontecorvo-Maki-Nakagawa-Sakata matrix, because this matrix, for the Majorana neutrino,   includes undetermined factors (Majorana phases). Obviously, if a value of the half-life, or a value of the effective neutrino mass, is assumed, the other can be determined.

\begin{figure}[t] 
\includegraphics[width=0.75\columnwidth]{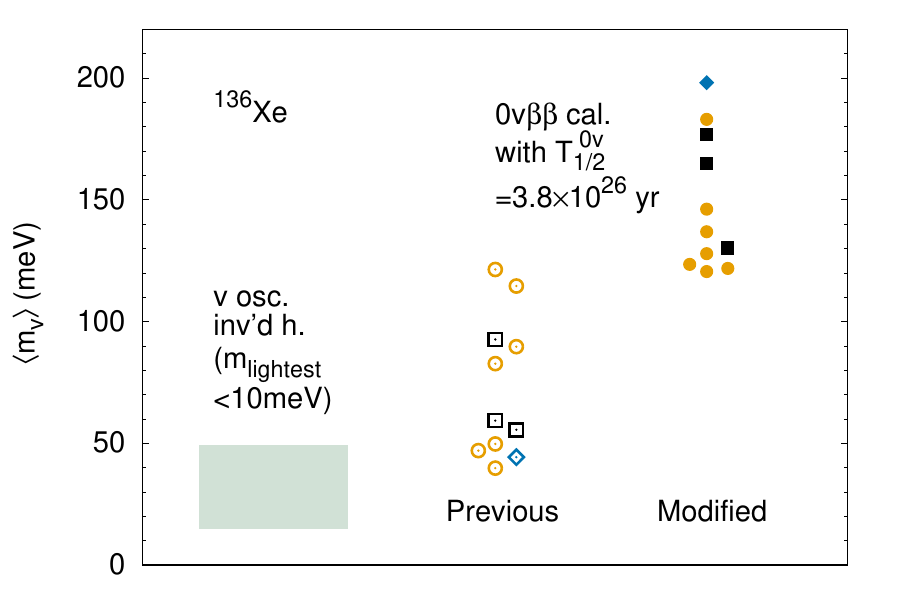}
\vspace{-10pt}
\caption{\label{fig:mv_o_m} \baselineskip=13pt 
Effective neutrino mass $\langle m_\nu\rangle$ from $^{136}$Xe$\rightarrow ^{136}$Ba. Those obtained from $g_A^2 M_{0\nu}$ of Fig.~\ref{fig:gA2M0v} are shown, where the current experimental lower limit of $T_{1/2}^{0\nu}$ = 3.8$\times 10^{26}$ yr is used. The region allowed by the neutrino oscillation data is also shown.  For this, the inverted hierarchy (ordering) of the neutrino eigen-masses and the lightest neutrino eigen-mass $<$ 10 meV are assumed. For the symbols, see the caption of Fig.~\ref{fig:gA2M0v}.}
\end{figure}

Figure \ref{fig:mv_o_m} illustrates $\langle m_\nu \rangle$ corresponding to $g_A^2 M_{0\nu}$ of Fig.~\ref{fig:gA2M0v} with the current experimental lower limit of $T_{1/2}^{0\nu}$ = 3.8$\times 10^{26}$ yr for $^{136}$Xe \cite{Abe25}. The neutrino oscillation experiments give a constraint on $\langle m_\nu\rangle$. The allowed region of $\langle m_\nu\rangle$ is also drawn in Fig.~\ref{fig:mv_o_m}, where  the inverted hierarchy (ordering) of the neutrino eigen-masses and the lightest eigen-mass $<$ 10 meV are assumed. The difference between the previous and the modified NMEs is significant. If the previous one is correct, the current experiment has the possibility to find the $0\nu\beta\beta$ decay. If the modified one is correct, there is no possibility to find it yet. 

\begin{figure}[t] 
\includegraphics[width=0.75\columnwidth]{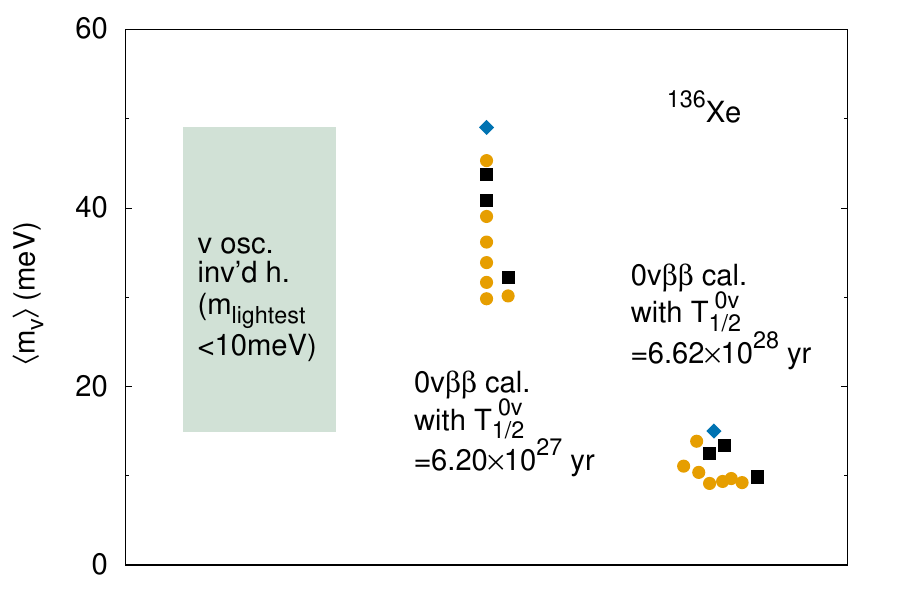}
\vspace{-10pt}
\caption{\label{fig:mv_16_174} \baselineskip=13pt 
The same as Fig.~\ref{fig:mv_o_m} but for different $T_{1/2}^{0\nu}$. Both distributions of $\langle m_\nu\rangle$ by the NME calculation are obtained from the modified NMEs.}
\end{figure}

\begin{figure}[t] 
\includegraphics[width=0.5\columnwidth]{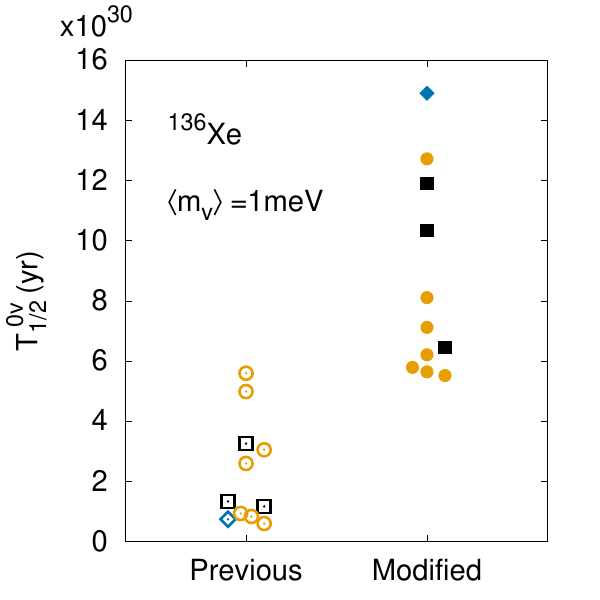}
\vspace{-10pt}
\caption{\label{fig:halflife} \baselineskip=13pt 
$T_{1/2}^{0\nu} (^{136}\mathrm{Xe})$ with assumed $\langle m_\nu \rangle$ = 1 meV and the values of $g_A^2 M_{0\nu}$ of Fig.~\ref{fig:gA2M0v}. For the symbols, see the caption of Fig.~\ref{fig:gA2M0v}.}
\end{figure}

The two distributions of $\langle m_\nu\rangle$ obtained from the NMEs have  comparable spreading. In fact, this implies that the uncertainty is reduced in the modified calculation. Due to the reciprocal relation of $\langle m_\nu\rangle$ and $M_{0\nu}$, the distribution of higher $\langle m_\nu\rangle$ is broader if the same NMEs are used. This is seen by using longer $T_{1/2}^{0\nu}$, as shown in Fig.~\ref{fig:mv_16_174}. $T_{1/2}^{0\nu}$ = 6.20$\times 10^{27}$ yr was chosen to put the highest calculated point at the upper edge of the region allowed by the neutrino oscillation data, and $T_{1/2}^{0\nu}$ = 6.62$\times 10^{28}$ yr was chosen to put the highest calculated point at the lower edge. With the former $T_{1/2}^{0\nu}$, the calculated $\langle m_\nu\rangle$ are included in the allowed region. The $0\nu\beta\beta$ decay can be found if the two assumptions for the allowed region are correct. If the experimental lower limit of $T_{1/2}^{0\nu}$ reached 6.62$\times 10^{28}$ yr, the possibility of the inverted hierarchy is excluded. 
As the lower limit of $T_{1/2}^{0\nu}$ becomes longer, the influence of the uncertainty of the NMEs on $\langle m_\nu\rangle$ is reduced. 
Figure \ref{fig:halflife} illustrates $T_{1/2}^{0\nu}$ with an  assumed $\langle m_\nu \rangle$ = 1 meV; this is in the normal mass hierarchy region. The modification shifts up $T_{1/2}^{0\nu}$, and the distribution is broadened. If $\langle m_\nu\rangle$ is one order of magnitude smaller, $T_{1/2}^{0\nu}$ is two orders of magnitude longer. 

\begin{table} 
\caption{\label{tab:NME0v2v} The calculated GT and Fermi NMEs of the $0\nu\beta\beta$ and the $2\nu\beta\beta$ decays of $^{76}$Ge and $^{136}$Xe. The nuclear-structure calculations were performed with the SkM$^\ast$ interaction. Leading, vc, and RS2bc stand for the leading order, the vertex correction, and the RS two-body current terms, respectively. Sum is called the perturbed NME. The sign is chosen so that the GT leading terms are positive. The results for $^{136}$Xe were taken from Ref.~\cite{Ter25b}. }
\begin{ruledtabular}
\begin{tabular}{ccccccccc}
& \multicolumn{4}{c}{$^{136}$Xe$\,\rightarrow^{136}$Ba} & \multicolumn{4}{c}{$^{76}$Ge$\,\rightarrow^{76}$Se}\\
\cline{2-5} \cline{6-9}\\[-10pt]
 & \multicolumn{2}{c}{$0\nu\beta\beta$ NME} & \multicolumn{2}{c}{$2\nu\beta\beta$ NME} 
 & \multicolumn{2}{c}{$0\nu\beta\beta$ NME} & \multicolumn{2}{c}{$2\nu\beta\beta$ NME ($\times 10^{-2}$) } \\
\cline{2-3} \cline{4-5} \cline{6-7} \cline{8-9}\\[-11pt]
 & GT & Fermi & GT & Fermi & GT & Fermi & GT & Fermi \\
\hline \\[-10pt]
Leading & $\;\;\:$3.095 &      $-$0.467 & $\;\;\:$0.102 &      $-$0.002 & $\;\;\:$2.278 & $-$0.368 & $\;\;\:$7.592 & $\;\;\:$0.016 \\
vc      & $\;\;\:$1.332 &      $-$0.984 & $\;\;\:$0.055 &      $-$0.033 & $\;\;\:$0.565 & $-$0.373 & $-$1.011 & $\;\;\:$0.403\\
RS2bc     &      $-$2.731 & $\;\;\:$1.758 &      $-$0.192 & $\;\;\:$0.030 & $-$0.566 & $\;\;\:$0.370 & $\;\;\:$1.873 & $-$0.940\\
Sum     & $\;\;\:$1.696 & $\;\;\:$0.307 &      $-$0.035 &      $-$0.005 & $\;\;\:$2.277 & $-$0.373 & $\;\;\:$8.454 & $-$0.519\\
\end{tabular}
\end{ruledtabular}
\end{table}
\begin{table} 
\caption{\label{tab:gAeff0v2v} Obtained $g_A^\mathrm{eff}$ for four definitions. The top one is for the $0\nu\beta\beta$ NME, and the others are for the $2\nu\beta\beta$ NME. The bottom one, $g_{A,2\nu}^\mathrm{eff}(\mathrm{pt;exp})$, is the $g_{A,2\nu}^\mathrm{eff}$ that is used with the perturbed NME components and reproduces the experimental $2\nu\beta\beta$ half-life. Those $g_A^\mathrm{eff}$ for $^{136}$Xe and $^{76}$Ge are shown. The former result was taken from Ref.~\cite{Ter25b}.}
\begin{ruledtabular}
\begin{tabular}{ccc}
 Specification of $g_A^\mathrm{eff}$ & $^{136}$Xe$\rightarrow ^{136}$Ba & $^{76}$Ge$\rightarrow ^{76}$Se\\
\hline \\[-10pt]
$g_{A,0\nu}^\mathrm{eff}$(ld;pt)  & 0.796 &  1.267 \\[+5pt]
$g_{A,2\nu}^\mathrm{eff}$(ld;pt)  & 0.696 &  1.363 \\[+5pt]
$g_{A,2\nu}^\mathrm{eff}$(ld;exp) & 0.422 &  1.177 \\[+6pt]
$g_{A,2\nu}^\mathrm{eff}$(pt;exp) & 0.806 &  1.087 \\
\end{tabular}
\end{ruledtabular}
\end{table}

\begin{figure}[t] 
\begin{minipage}[t]{1.0\columnwidth}
\includegraphics[width=0.75\columnwidth]{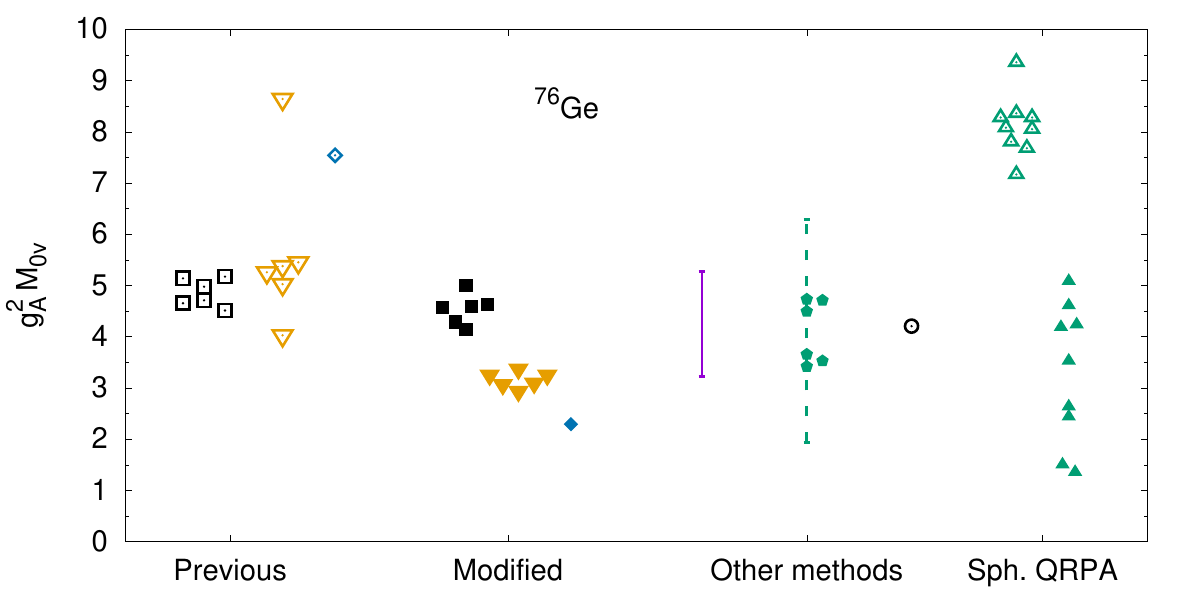}
\includegraphics[width=0.35\columnwidth]{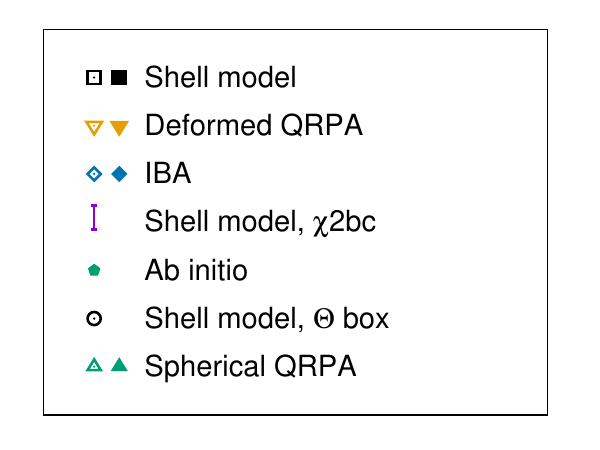}
\vspace{-10pt}
\caption{\label{fig:gA2M0v_ge76} \baselineskip=13pt 
The calculated $g_A^2 M_{0\nu}$ for $^{76}$Ge. Those results of the previous and modified results are obtained in the same manner as for the case of $^{136}$Xe. Each of these two groups includes the results of the Shell Model \cite{Bro15,Men18,Cor20b}, the deformed QRPA (Refs.~\cite{Mus13,Fan18} and  our present calculation), and the IBA \cite{Bar15}. The ground state of $^{76}$Ge is deformed. Other methods are the results of the Shell Model with two body currents of the $\chi$EFT ($\chi$2bc) \cite{Men11} type, the ab initio method \cite{Bel24}, and the Shell Model with the Suzuki-Okamoto method ($\Theta$ box) \cite{Cor20b}. For these results, $g_A^\mathrm{bare}$ is used. The dashed line of ab initio indicates the 68\% confidence line. The results shown in the column of Sph.~QRPA are obtained according to the previous (open symbols) and the modified (filled symbols) methods with the spherical QRPA \cite{Bro15,Eng14,Sim18b,Hyv15}. Those symbols are defined in the separate panel.}
\end{minipage}
\end{figure}

\section{\label{sec:ge76}The decay of $^{\mathbf{76}}$G\lowercase{e} }
We discuss the $\beta\beta$ decay instance of $^{76}\mathrm{Ge}$ $\rightarrow$ $^{76}$Se. The components of the $\beta\beta$ NME for $^{76}$Ge are shown in Table \ref{tab:NME0v2v} together with those for $^{136}$Xe \cite{Ter25b}. The most remarkable feature of $^{76}$Ge is that the vertex correction and the RS two-body current terms in the $0\nu\beta\beta$ NME cancel each other, so that the NMEs  converge at the leading order. The absolute values of these correction terms for the GT component are much smaller than the leading terms for both the decay modes. This is also the difference respect to the case of  $^{136}$Xe. The   Fermi $2\nu\beta\beta$ NME vanishes exactly if the system has the isospin symmetry. 
This cancellation is partial for $^{76}$Ge; still, the perturbed (sum) Fermi $2\nu\beta\beta$ NME is less than 10\% of the GT term. The expected cancellation is approximately satisfied. The calculation of $^{136}$Xe satisfies this cancellation well. It is seen from Table \ref{tab:NME0v2v} that the nuclear mass dependence of the NME components is significant. One may be concerned whether the calculated $2\nu\beta\beta$ half-life reproduces the experimental value without $g_A^\mathrm{eff}$. The calculated value with $g_A^\mathrm{bare}$ is 1.40$\times 10^{21}$ yr, while the current experimental value is 1.88$\times 10^{21}$ yr \cite{Bar19}; the difference is approximately 25\%.   This agreement is 
remarkable. 
Table \ref{tab:gAeff0v2v} shows values  of $g_A^\mathrm{eff}$ for $^{136}$Xe and $^{76}$Ge. The reduction of $g_A^\mathrm{eff}$ supplements the insufficient many-body correlations in the NME components. Any $g_A^\mathrm{eff}$ for $^{76}$Ge are closer to $g_A^\mathrm{bare}$ than those for $^{136}$Xe. This result is consistent with the convergence of the NME for $^{76}$Ge. The ratio of $g_{A,0\nu}^\mathrm{eff}(\mathrm{ld;pt})$ to $g_{A,2\nu}^\mathrm{eff}(\mathrm{ld;pt})$ is equal to 0.930, which is used for the renormalization factor in Eq.~(\ref{eq:gA0veff_est}). 
The estimated $g_{A,0\nu}^\mathrm{eff}(\mathrm{ld;est})$ is equal to 1.094.   
The bare value of $M_{0\nu}^{(0)}$ [Eq.~(\ref{eq:gA2M0v_previous})] is obtained to be 2.508. If $g_{A,0\nu}^\mathrm{eff}(\mathrm{ld;est})$ is used for $M_{0\nu}^{(0)}$, this NME is 2.586; see Eq.~(\ref{eq:gA2M0v_modified}). Thus, the bare $M_{0\nu}^{(0)}$ value is already very close to the one by our modification method.

The $g_A^2 M_{0\nu}$ calculated by several groups and our modification method are summarized in Fig.~\ref{fig:gA2M0v_ge76}. The previous and the modified results are obtained in the same manner as those for $^{136}$Xe, that is using  Eqs.~(\ref{eq:gA2M0v_previous}) and (\ref{eq:gA2M0v_modified}), respectively. 
The results obtained by other methods include the perturbed NME of Refs.~\cite{Men11,Bel24,Cor20b} with $g_A^\mathrm{bare}$. For $^{76}$Ge, we show separately the results of the deformed QRPA (those of Refs.~\cite{Mus13,Fan18} and ours), and the spherical QRPA \cite{Bro15,Eng14,Sim18b,Hyv15} because at least the ground state of $^{76}$Ge has a quadrupole deformation. The spherical QRPA calculations assume the spherical shape of the nuclear density distribution. The results of the deformed QRPA are included in the previous and modified ones in Fig.~\ref{fig:gA2M0v_ge76}. The authors of Refs.~\cite{Cor19,Cor20b} have calculated the $0\nu\beta\beta$ and $2\nu\beta\beta$ NMEs using the Shell Model and their own method to perturb the NMEs. Their results can give the renormalization factor for $g_{A,0\nu}^\mathrm{eff}(\mathrm{pt;est})$ in Eq.~(\ref{eq:gA0veff_est}), which is equal to 1.517. This value was used for the modification of the Shell Model results. For the QRPA modifications, our renormalization factor was used, as well  as for the modification of the IBA results.

The significant difference between $^{76}$Ge$\,$-$^{76}$Se and $^{136}$Xe$\,$-$^{136}$Ba is the deformation. The latter nuclei are spherical. The deformation effects on $^{76}$Ge and $^{76}$Se have been investigated intensively in spectroscopic studies. Relatively recent ones are found in Refs.~\cite{Guo07,Toh13,Nik14,Nom22,Aya23}. 
Two conclusions are established. 
First, the ground state of $^{76}$Ge has a quadrupole deformation, which is $\beta \simeq 0.2$ \cite{Nik14,Nom22}. Reference \cite{Aya23} reports the quadrupole moment of $\langle Q^2 \rangle = 3000\ e^2\mathrm{fm}^2$. The quadrupole deformation is consistent with the experimental data of the rotational-band spectrum and the strong $E2$ transitions between the member states \cite{nndc26}. 
Second, the potential energy surface is substantially shallow with respect to the $\gamma$ degree of freedom for both $^{76}$Ge and $^{76}$Se. The rotational spectrum of $^{76}$Se is not as clear as that of $^{76}$Ge \cite{nndc26}. According to our HFB calculation with SkM$^\ast$ and that of Ref.~\cite{Mus13}, the ground state of $^{76}$Se is spherical, while the potential energy surface of Ref.~\cite{Nom22} shows a clear quadrupole deformation. The $\gamma$ of the energy minimum in the shallow potential energy surface is not yet established. This uncertainty is related to the historical question of whether there is a rigid  triaxial asymmetric deformation.
 
As Fig.~\ref{fig:gA2M0v_ge76} shows, the previous values of $g_A^2 M_{0\nu}$ given by  the deformed QRPA calculations are distributed in the range of 4.0$-$8.6. This range is reduced and lowered to 2.9$-$3.4. The reduction rate of the dispersion is 89\%; this is the most significant 
 reduction obtained with our modification method. 
 The spherical QRPA approach yields the previous values of $g_A^2 M_{0\nu}$ in the interval of 7.2$-$9.4 and the modified ones of 1.4$-$5.1. The dispersion is broadened by our modification method. The calibration of the calculation using the experimental data does not narrow the dispersion of these  theoretical results, implying  that the $0\nu\beta\beta$ and $2\nu\beta\beta$ NMEs are not correlated in the spherical QRPA approach. 
Since $^{76}$Ge is deformed, the reduction of the dispersion of deformed-QRPA NMEs is appropriate, and the  $0\nu\beta\beta$ and the $2\nu\beta\beta$ NMEs are correlated. 

\begin{figure}[] 
\includegraphics[width=0.75\columnwidth]{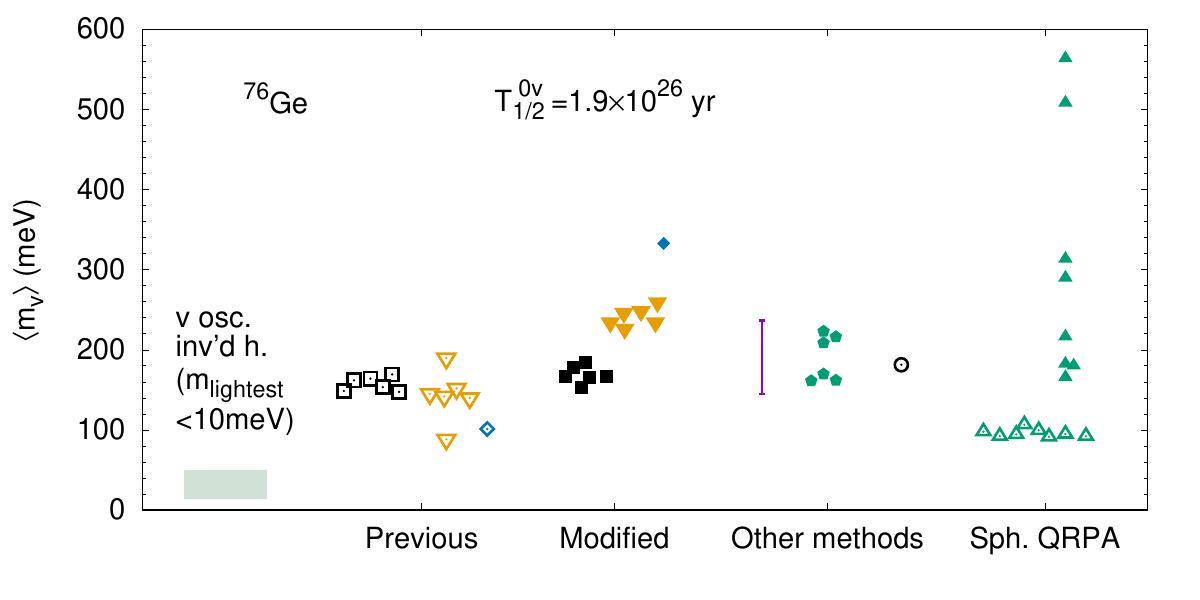}\vspace{-10pt}
\caption{\label{fig:mv_ge76} \baselineskip=13pt 
Effective neutrino mass $\langle m_\nu\rangle$. Those of the nuclear calculations for $^{76}$Ge$\rightarrow ^{76}$Se were obtained from $g_A^2 M_{0\nu}$ of Fig.~\ref{fig:gA2M0v_ge76} and the current experimental lower limit of the $0\nu\beta\beta$ half-life $T_{1/2}^{0\nu}$ \cite{Ach26}. The region by the neutrino oscillation data is the same as that of Fig.~\ref{fig:mv_o_m}. For the symbols, see Fig.~\ref{fig:gA2M0v_ge76}.}
\end{figure}

\begin{figure}[] 
\includegraphics[width=0.75\columnwidth]{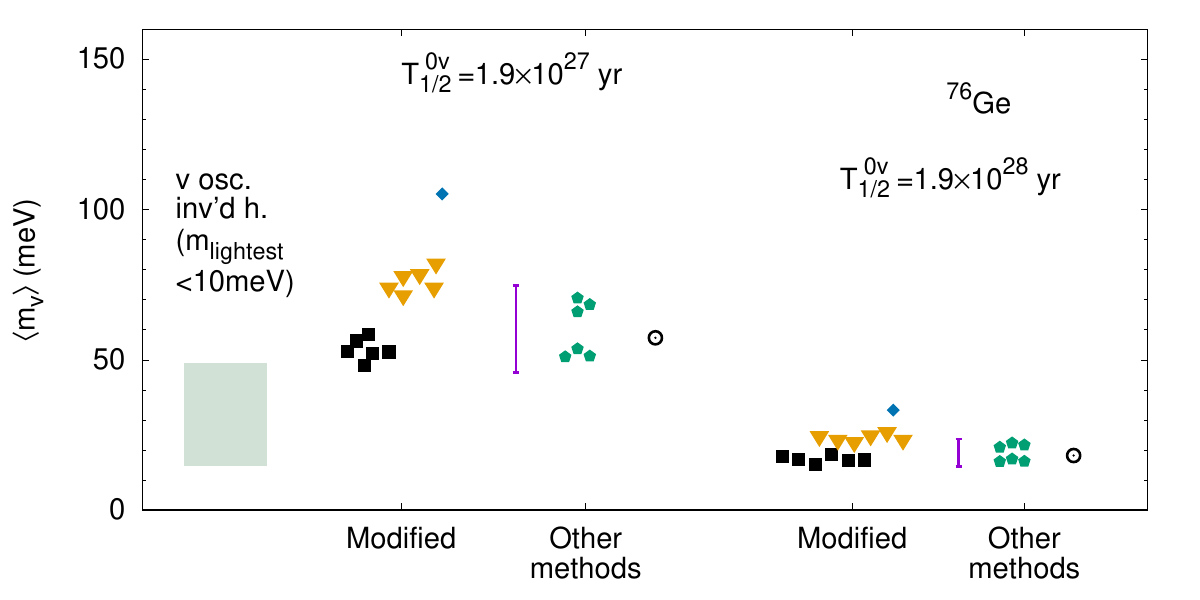}\vspace{-10pt}
\caption{\label{fig:mv_10_76Ge} \baselineskip=13pt 
Effective neutrino mass $\langle m_\nu\rangle$ obtained by the same method as Fig.~\ref{fig:mv_ge76} but for assumed longer half-lives. For the symbols, see Fig.~\ref{fig:gA2M0v_ge76}.}
\end{figure}

\begin{figure}[] 
\includegraphics[width=0.5\columnwidth]{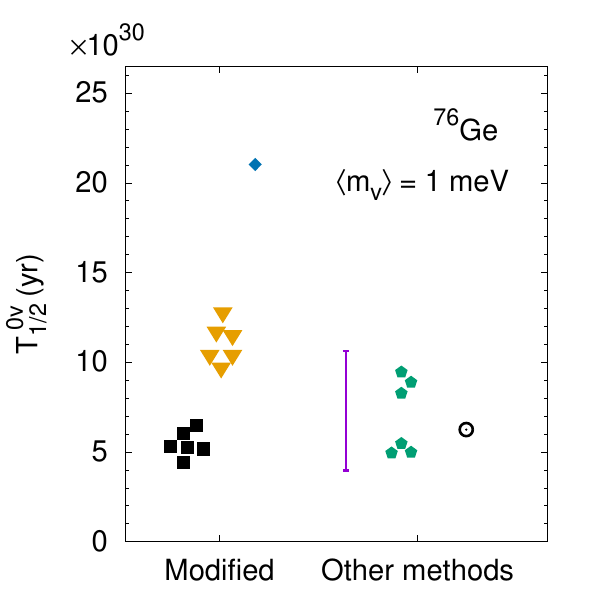}\vspace{-10pt}
\caption{\label{fig:hl_ge76} \baselineskip=13pt 
Half-life of the $0\nu\beta\beta$ decay of $^{76}$Ge with a tentative assumption of $\langle m_\nu\rangle = 1$ meV. For the symbols, see Fig.~\ref{fig:gA2M0v_ge76}.}
\end{figure}

It is shown by Fig.~\ref{fig:gA2M0v_ge76} that $g_{A,0\nu}^2 M_{0\nu}$ of the Shell Model  and those of other methods have large overlaps. The modified results of the deformed QRPA are close to those two groups.  The values of $g_{A,0\nu}^2 M_{0\nu}$ for $^{76}$Ge are in the range of 2.3$-$5.0, which does not include the previous ones and the spherical-QRPA results. The range before our modification is 4.0$-$8.6. The mean value is lowered by 2.7, and the reduction rate of the dispersion is 41\%.  We mention an uncertainty of the QRPA, which is not a good approximation for large dynamical fluctuations. Approximations with extensions of the QRPA could reduce the NME due to the $\gamma$ softness. 
Figure \ref{fig:mv_ge76} illustrates the values of $\langle m_\nu\rangle$ obtained from the nuclear-structure  calculations with the current experimental lower limit of the half-life, and Fig.~\ref{fig:mv_10_76Ge} shows $\langle m_\nu\rangle$ with longer half-lives. $T_{1/2}^{0\nu}$ of order of 10$^{28}$ yr is crucial for covering the inverted hierarchy region. Finally, the half-life for $\langle m_\nu\rangle = 1 $ meV is drawn in Fig.~\ref{fig:hl_ge76}. The majority of the calculated results are in the range of (5$-$13)$\times 10^{30}$ yr. Its maximum-to-minimum ratio is 2.5. 

\section{\label{sec:130Te_110Pd}The decays of $^{\mathbf{130}}$T\lowercase{e} and $^{\mathbf{110}}$P\lowercase{d}}

\begin{table} 
\caption{ \label{tab:beta_130Te_110Pd} Axially symmetric quadrupole deformation parameter $\beta$ of the HFB ground states of the initial and final nuclei of the two decay instances discussed in Sec.~\ref{sec:130Te_110Pd}. }
\centering
\begin{minipage}{0.3\textwidth}
\begin{ruledtabular}
\begin{tabular}{cc}
Nucleus & $\beta$ \\
\hline \\[-10pt]
$^{130}$Te & 0    \\
$^{130}$Xe & 0.11 \\
$^{110}$Pd & 0.25 \\
$^{110}$Cd & 0.16 \\
\end{tabular}
\end{ruledtabular}
\end{minipage}
\end{table}

\begin{table} 
\caption{\label{tab:NME0v2v_130Te} The calculated GT and Fermi NMEs of the $0\nu\beta\beta$ and the $2\nu\beta\beta$ decays of $^{130}$Te and $^{110}$Pd. The nuclear-structure calculations were performed with the SkM$^\ast$ interaction. See the caption of Table \ref{tab:NME0v2v} for the items in the leftmost column. }
\begin{ruledtabular}
\begin{tabular}{ccccccccc}
& \multicolumn{4}{c}{$^{130}$Te$\,\rightarrow^{130}$Xe} & \multicolumn{4}{c}{$^{110}$Pd$\,\rightarrow^{110}$Cd}\\
\cline{2-5} \cline{6-9}\\[-10pt]
 & \multicolumn{2}{c}{$0\nu\beta\beta$ NME} & \multicolumn{2}{c}{$2\nu\beta\beta$ NME} 
 & \multicolumn{2}{c}{$0\nu\beta\beta$ NME} & \multicolumn{2}{c}{$2\nu\beta\beta$ NME  } \\
\cline{2-3} \cline{4-5} \cline{6-7} \cline{8-9}\\[-11pt]
 & GT & Fermi & GT & Fermi & GT & Fermi & GT & Fermi ($\times 10^{-2}$) \\
\hline \\[-10pt]
Leading & $\;\;\:$3.613 &      $-$0.709 & $\;\;\:$0.132 & $-$3$\times 10^{-4}$ & $\;\;\:$2.486 &      $-$0.571 & $\;\;\:$0.270 &      $-$0.105 \\
vc      & $\;\;\:$1.248 &      $-$0.899 &      $-$0.014 &      $-$0.021        & $\;\;\:$0.769 &      $-$0.508 & $-$0.033      &      $-$0.767 \\
RS2bc   &      $-$2.155 & $\;\;\:$1.443 & $\;\;\:$0.022 & $\;\;\:$0.023        &      $-$1.429 & $\;\;\:$0.900 & $\;\;\:$0.272 & $\;\;\:$1.48  \\
Sum     & $\;\;\:$2.706 &      $-$0.165 & $\;\;\:$0.139 & $\;\;\:$0.001        & $\;\;\:$1.827 &      $-$0.179 & $\;\;\:$0.510 & $\;\;\:$0.61  \\
\end{tabular}
\end{ruledtabular}
\end{table}

\begin{table} 
\caption{\label{tab:gAeff0v2v_130Te} Obtained $g_A^\mathrm{eff}$ for $^{130}$Te and $^{110}$Pd. The definitions of the top two $g_A^\mathrm{eff}$ are given by Eqs.~(\ref{eq:cndn_gaeff0v}) and (\ref{eq:cndn_gaeff2v}). The $g_{A,2\nu}^\mathrm{eff}(\mathrm{ld;exp})$ and $g_{A,2\nu}^\mathrm{eff}(\mathrm{pt;exp})$ are defined in Sec.~\ref{sec:method} and the caption of Table \ref{tab:gAeff0v2v}, respectively. Their values are noted if the experimental $T_{1/2}^{2\nu}$ is available. We used the half-life value of 7.91$\times 10^{20}$ yr for $^{130}$Te \cite{Bar19} to derive 
$g_{A,2\nu}^\mathrm{eff}$(ld;exp) and $g_{A,2\nu}^\mathrm{eff}$(pt;exp). }
\begin{ruledtabular}
\begin{tabular}{ccc}
 Specification of $g_A^\mathrm{eff}$ & $^{130}$Te$\rightarrow ^{130}$Xe & $^{110}$Pd$\rightarrow ^{110}$Cd\\
\hline \\[-10pt]
$g_{A,0\nu}^\mathrm{eff}$(ld;pt)  & 1.028 &  1.014 \\[+5pt]
$g_{A,2\nu}^\mathrm{eff}$(ld;pt)  & 1.302 &  1.404 \\[+5pt]
$g_{A,2\nu}^\mathrm{eff}$(ld;exp) & 0.465 &   \\[+6pt]
$g_{A,2\nu}^\mathrm{eff}$(pt;exp) & 0.462 &   \\
\end{tabular}
\end{ruledtabular}
\end{table}

We investigate two more decay instances in $A = 100{-}130$ region. That is $^{130}$Te$\rightarrow$$^{130}$Xe and $^{110}$Pd$\rightarrow$$^{110}$Cd. 
The leading-order NMEs of these decay instances were calculated by our QRPA approach previously \cite{Ter20}. 
First of all, we show in Table \ref{tab:beta_130Te_110Pd} the axially symmetric quadrupole deformation parameter $\beta$ obtained by the HFB calculation for the ground states. All of the four nuclei except for $^{130}$Te are deformed. 

Table \ref{tab:NME0v2v_130Te} shows the NME components of the $0\nu\beta\beta$ and $2\nu\beta\beta$ decays of $^{130}$Te and $^{110}$Pd. The mass number difference between $^{130}$Te and $^{136}$Xe is small. Thus, it is worthy of comparing these two instances; for $^{136}$Xe see Table \ref{tab:NME0v2v}. The leading and the correction terms are similar between the two instances for both the $0\nu\beta\beta$ and $2\nu\beta\beta$ NMEs. The order in the absolute value and the signs are the same for the $0\nu\beta
\beta$ NME. The perturbed GT  $0\nu\beta\beta$ NME (sum) of $^{130}$Te is larger than that of $^{136}$Xe by 60\%. The difference in the leading GT $0\nu\beta\beta$ NME between the two instances is 15\%. The partial cancellation between the leading and correction terms makes the difference in the perturbed term between the two instances relatively large. The signs of the perturbed Fermi components are different. 

The leading GT component of the $2\nu\beta\beta$ NME is similar between $^{130}$Te and $^{136}$Xe, but the two GT RS2bc are different by a factor of 9. The two correction terms of the GT $2\nu\beta\beta$ NME of $^{130}$Te cancel appreciably, and the leading and the perturbed GT components are close. The sign of the perturbed GT $2\nu\beta\beta$ NME is different between the two instances. For the Fermi component of the $2\nu\beta\beta$ NME, the cancellation of the vc and the RS2bc is common, and the perturbed Fermi component is much smaller than the perturbed GT component for both the instances. The isospin symmetry is approximately satisfied. Thus, there are the overall similarities and the differences in details. 

Comparing $^{110}$Pd and $^{130}$Te, one sees the common point that the vc and the RS2bc terms always have the opposite signs. The overall trend of the leading and the correction terms is similar. The approximate isospin symmetry is also well satisfied by the Fermi component of the $2\nu\beta\beta$ NME of $^{110}$Pd. 

The ratio of $g_{A,0\nu}^\mathrm{eff}$(ld;pt) to $g_{A,2\nu}^\mathrm{eff}$(ld;pt) is crucial for our modification method. From Table \ref{tab:gAeff0v2v_130Te}, the ratio is found to be 0.790 ($^{130}$Te) and 0.722 ($^{110}$Pd). Their deviations from one are small. The peak neutrino momentum of the $0\nu\beta\beta$ decay is $\sim$100 MeV/c, while that of the $2\nu\beta\beta$ decay is at most 2$-$3 MeV/c. The ratio of $g_{A,0\nu}^\mathrm{eff}$(ld;pt) to $g_{A,2\nu}^\mathrm{eff}$(ld;pt) is approximately one compared to the ratio of the neutrino momenta. Our calculations indicate that the influence of the neutrino momentum in the $0\nu\beta\beta$ decay on $g_{A,0\nu}^\mathrm{eff}$ is generally small. 
The $g_{A,2\nu}^\mathrm{eff}$(pt;exp) and $g_{A,2\nu}^\mathrm{eff}$(ld;exp) of $^{130}$Te are close, reflecting that the leading and the perturbed GT components are close. Those small $g_{A,2\nu}^\mathrm{eff}$ values of $\approx$0.46 imply that the experimental $T_{1/2}^{2\nu}$ is not reproduced if $g_A^\mathrm{bare}$ is used. This is in contrast to our $^{76}$Ge calculation. For $^{130}$Te, the higher-order corrections are anticipated for the convergence of the NME if the NME with $g_A^\mathrm{bare}$ should reproduce the experimental half-life. Since the ratio of the theoretical $g_{A,0\nu}^\mathrm{eff}$ and $g_{A,2\nu}^\mathrm{eff}$ is approximately converged at the first order, we use this advantage to avoid the difficulty in the convergence of the $0\nu\beta\beta$ NME. The higher-order corrections, i.e. the many-body correlations, are included in the phenomenological $g_{A,2\nu}^\mathrm{eff}$ non-perturbatively. 
The effect of the deformation on the $\beta\beta$ NME is not  obvious in the results. The experimental $T_{1/2}^\mathrm{2\nu}$ of $^{110}$Pd is not yet measured.

\begin{figure}[t] 
\includegraphics[width=0.5\columnwidth]{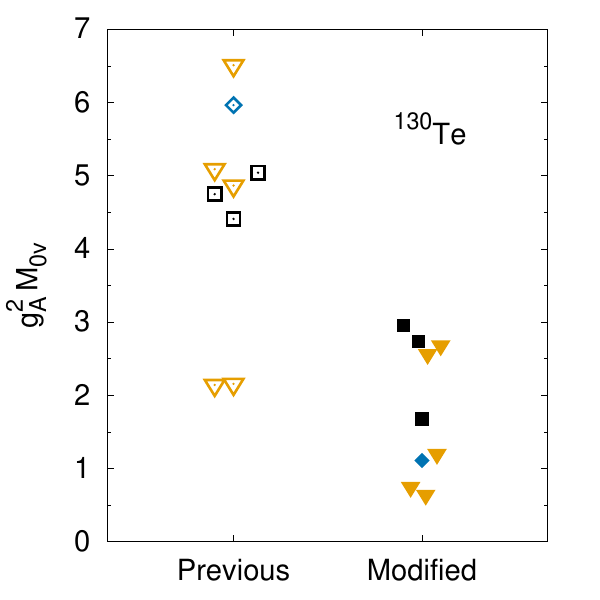}
\vspace{-10pt}
\caption{\label{fig:gA2M0v_130Te} \baselineskip=13pt The previous (usual) and the modified $g_A^2 M_{0\nu}$ for $^{130}$Te. The squares, triangles, and diamond show the results of the shell model, deformed QRPA, and IBA. The results of the spherical QRPA are not included. The references are the same as those used for the discussions of $^{76}$Ge except for those spherical QRPA, the Horoi et al.'s shell model calculation, and the ab initio calculation.}
\end{figure}

\begin{figure}[t] 
\includegraphics[width=0.67\columnwidth]{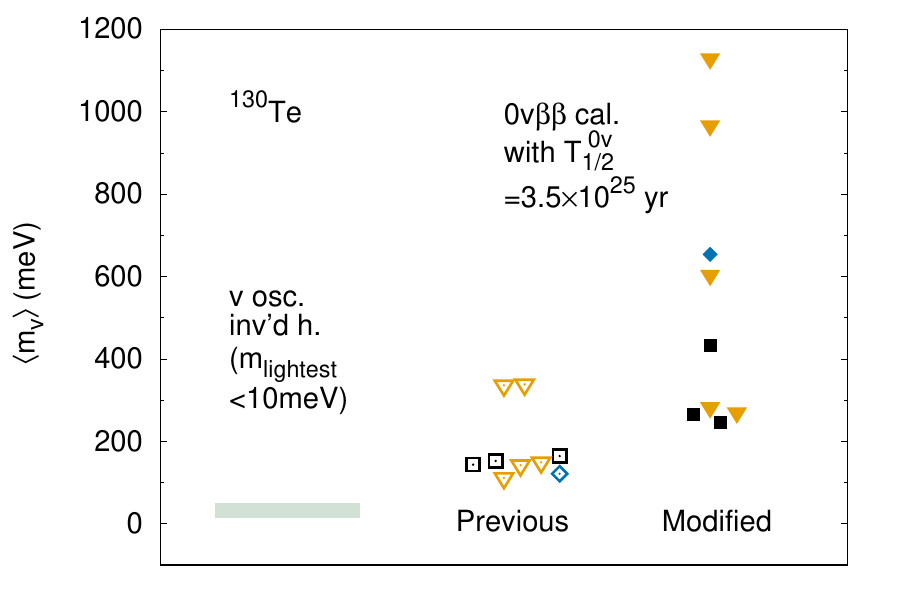}
\vspace{-10pt}
\caption{\label{fig:mv_om_130Te} \baselineskip=13pt Effective neutrino mass $\langle m_\nu\rangle$. The rectangle shows the region allowed by the neutrino oscillation data (the inverted ordering and $m_\mathrm{lightest}$ $<$ 10 meV). The half-life used for the $\langle m_\nu\rangle$ calculation is the current experimental lower limit of 3.5$\times$$10^{25}$ yr for $^{130}$Te \cite{CUO25}. For the symbols, see the caption of Fig.~\ref{fig:gA2M0v_130Te}. }
\end{figure}

\begin{figure}[t] 
\includegraphics[width=0.67\columnwidth]{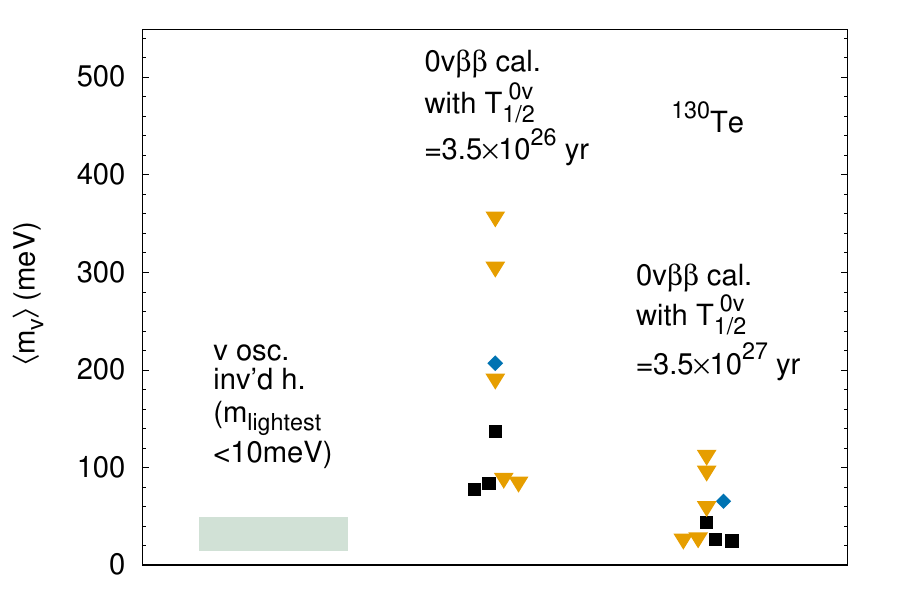}
\vspace{-10pt}
\caption{\label{fig:mv_10_100_130Te} \baselineskip=13pt The same plot as the modified group in Fig.~\ref{fig:mv_om_130Te} but for different assumed half-life values, which are those 10 times and 100 times longer than the current lower limit.  }
\end{figure}

\begin{figure}[t] 
\includegraphics[width=0.5\columnwidth]{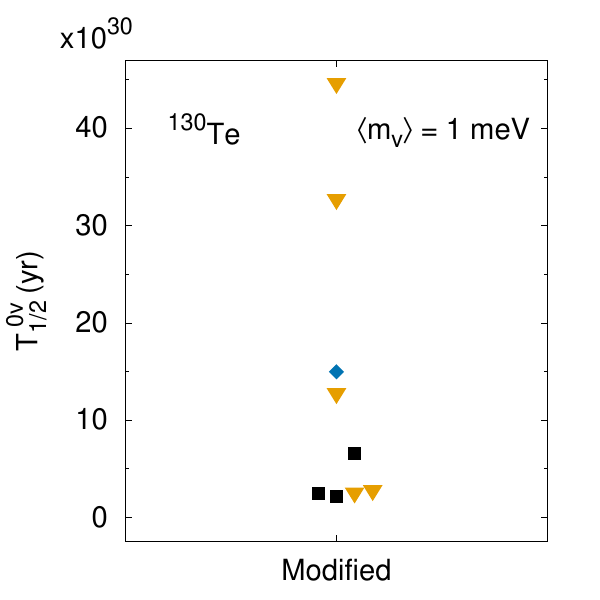}
\vspace{-10pt}
\caption{\label{fig:halflife_130Te} \baselineskip=13pt Calculated $T_{1/2}^{0\nu}$ of $^{130}$Te with assumed $\langle m_\nu\rangle$ = 1 meV. The modified $g_A^2 M_{0\nu}$ are used. For the symbols, see the caption of Fig.~\ref{fig:gA2M0v_130Te}.}
\end{figure}

\begin{figure}[t] 
\includegraphics[width=0.50\columnwidth]{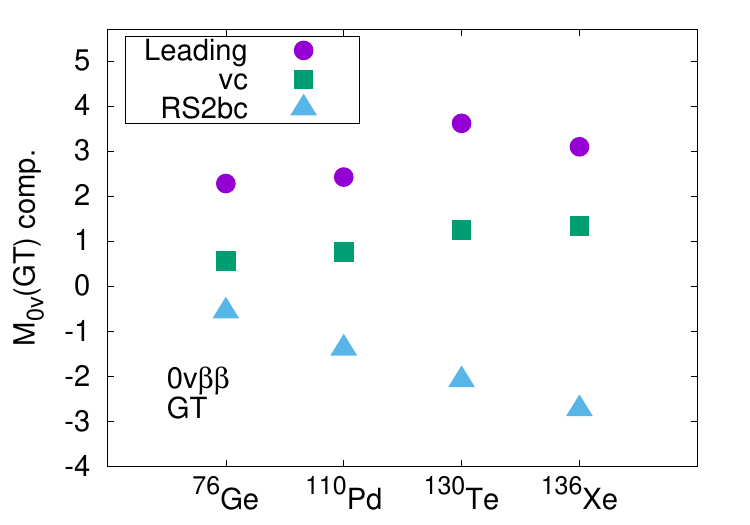}
\vspace{-10pt}
\caption{\label{fig:m0vgt_cmp} \baselineskip=13pt Decay instance dependence of the leading and the correction terms of the GT component of the $0\nu\beta\beta$ NME.}
\end{figure}

\begin{figure}[t] 
\includegraphics[width=0.50\columnwidth]{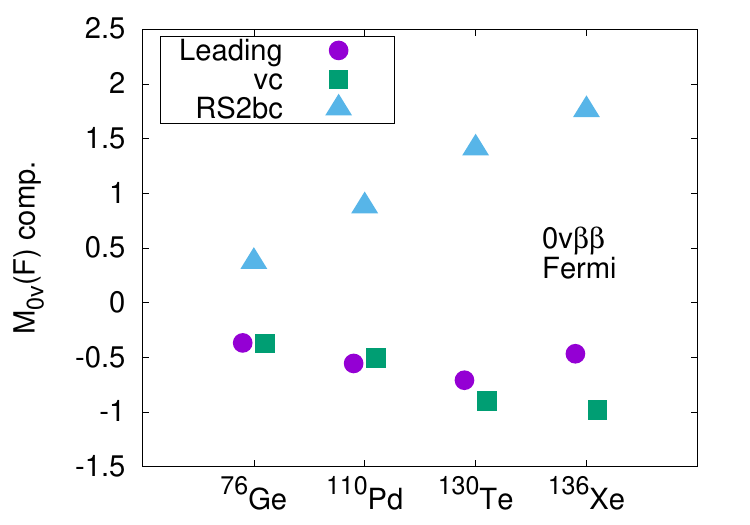}
\vspace{-10pt}
\caption{\label{fig:m0vfrm_cmp} \baselineskip=13pt The same as Fig.~\ref{fig:m0vgt_cmp} but for the Fermi component. }
\end{figure}

The previous and modified $g_A^2 M_{0\nu}$ for the decay of $^{130}$Te are shown in Fig.~\ref{fig:gA2M0v_130Te}. The overall trend is similar to that for $^{76}$Ge. The modified dispersion is 2.5, while the previous one is 4.4; the reduction rate is 43\%. The lowest modified one is lower than the lowest previous one by 1.5. 

Figure \ref{fig:mv_om_130Te} shows $\langle m_\nu \rangle$, including those obtained by the $^{130}$Te calculations. The overall features of the previous and the modified distributions are similar to those of the other nuclear decays. However, the calculated $\langle m_\nu \rangle$ are higher than the other cases mainly because the $T_{1/2}^{0\nu}$ used here is one order of magnitude shorter than those of the other instances. Figure \ref{fig:mv_10_100_130Te} depicts $\langle m_\nu \rangle$ for longer $T_{1/2}^{0\nu}$ values. As the corresponding figures of other instances, i.e.  Figs.~\ref{fig:mv_16_174} and \ref{fig:mv_10_76Ge}, the longer  $T_{1/2}^{0\nu}$ gives the lower and narrower distributions. With $T_{1/2}^{0\nu}$ of 3.5$\times$$10^{27}$ yr, the lower part of the distribution is in the region allowed by the neutrino oscillation data with the two assumptions.

Figure \ref{fig:halflife_130Te} illustrates $T_{1/2}^{0\nu}$ of $^{130}$Te obtained with the modified $g_A^2 M_{0\nu}$ and the assumed $\langle m_\nu \rangle$ = 1 meV. The range of the distribution is of the same order as those of $^{136}$Xe (Fig.~\ref{fig:halflife}) and $^{76}$Ge (Fig.~\ref{fig:hl_ge76}), but that of $^{130}$Te is higher than the others by a factor of 2$-$3. This difference reflects the nuclear properties.

Finally, we show the systematics of the leading and the correction terms of the four decay instances discussed so far in Figs.~\ref{fig:m0vgt_cmp} (GT) and \ref{fig:m0vfrm_cmp} (Fermi).  The most remarkable feature of these figures is that the RS2bc term has the strongest $A$ dependence. This feature can be understood diagramatically. The 2bc term involves three nucleons, and the one with no weak-interaction vertex is any of the other $A-2$ nucleons in the nucleus. Thus, the $A$ dependence of this term is different from other terms that involves two nucleons. This $A$ dependence explains why the correction term of $^{136}$Xe is large, and that of $^{76}$Ge cancels. 

\section{\label{sec:summary}Summary}
We have proposed a new method to calculate the $0\nu\beta\beta$ NMEs and demonstrated its effect using a set of different calculations for the decays of $^{136}$Xe, $^{130}$Te, $^{110}$Pd, and $^{76}$Ge. The principle is to apply $g_{A,2\nu}^\mathrm{eff}$ to the $0\nu\beta\beta$ calculation. This method is justified by the extended NME calculation with the perturbed transition operator, and by considering the role of the neutrino potential in $g_{A,0\nu}^\mathrm{eff}$.
 The important support for our method is that the ratio of $g_{A,0\nu}^\mathrm{eff}/g_{A,2\nu}^\mathrm{eff}$  nearly or already converges at the first-order perturbation. This similarity of the two theoretical values of $g_A^\mathrm{eff}$ is also seen by other approaches using the chiral two-body current and the $\Theta$ box method. The difference in the relevant momentum of the neutrino for the two decay modes is by two orders of magnitude. This enormous difference does not affect $g_{A,0\nu}^\mathrm{eff}$. In our approach, the many-body effects missing from the leading-order calculations are incorporated into $g_A^\mathrm{eff}$. Thus, we did not pursue the convergence of the perturbed NME components. Our method has two advantages. One is that the many-body effects can be  included in the NME non-perturbatively through the experimental data. The other one is that our method can be applied to the results obtained by  other groups using different nuclear-structure methods, as we have shown. 

The dispersion of $g_A^2 M_{0\nu}$ is dramatically reduced by our method for $^{136}$Xe compared to the previous results. For the other nuclei, the reduction rate is (40$-$50)\%. Another important outcome of our method is that $g_A^2 M_{0\nu}$ is lowered in many cases. 
The deformation effect was discussed in depth for $^{76}$Ge in the QRPA. If the deformation is ignored, the dispersion of the different results is not narrowed. 
We have discussed the implication of the modified NMEs for $\langle m_\nu\rangle$ and $T_{1/2}^{0\nu}$. The neutrino oscillation experiments give a constraint on $\langle m_\nu\rangle$. By combining this constraint and the modified NMEs, the prospects were obtained for the possibility to observe the $0\nu\beta\beta$ decay. The current experimental situation is approaching the inverted-hierarchy region of the neutrino eigen-masses. 

We have demonstrated the robustness of our modification method in $^{136}$Xe, $^{130}$Te, and $^{76}$Ge. The correction terms for $^{110}$Pd are qualitatively similar to those of the other nuclei. Our study explores a unique new direction toward a solution to the longstanding uncertainty problem of the $0\nu\beta\beta$ NME.

\begin{acknowledgments}  
This study was supported by the Czech Science Foundation (GA\v{C}R), project No. 24-10180S. The computation for this study was performed by Karolina (OPEN-33-78, OPEN-36-66), IT4Innovations supported by the Ministry of Education, Youth and Sports of the Czech Republic through the e-INFRA CZ (ID:90254); the computers of MetaCentrum provided by the e-INFRA CZ project (ID:90254), supported by the Ministry of Education, Youth and Sports of the Czech Republic; and Yukawa-21 at Yukawa Institute for Theoretical Physics, Kyoto University. O.C.~acknowledges the PIP 2081 (CONICET) and the PICT  40492 (ANPCyT). 
We thank ECT$^\ast$ (European Centre for Theoretical Studies in Nuclear Physics and Related Areas), INFN (Istituto Nazionale di Fisica Nucleare), EUROLABS funded by the European Union’s Horizon Europe Research and Innovation programme under grant agreement No 101057511, and EUSTIPEN (Europe-U.S.~Theory Institute for Physics with Exotic Nuclei), which is supported by FRIB Theory Alliance under DOE grant number DE-SC0013617, for support at the Workshop “The search for neutrinoless double-$\beta$ decay: experimental and theoretical challenges” during which this work has been developed. We thank the Institute for Nuclear Theory at the University of Washington for its hospitality during the INT-26-2a program, ``Recommended Values for Neutrinoless Double-Beta Decay Nuclear Matrix Elements,'' and the Department of Energy for partial support during our participation in this program.
\end{acknowledgments}


\bibliography{solution}
\end{document}